\documentclass[amssymb,twocolumn,aps,prl,showpacs]{revtex4}
\usepackage{amsmath,mathtools}
\usepackage{graphicx}
\usepackage{sidecap}
\usepackage{dcolumn}
\usepackage{bm}
\usepackage{hyperref}
\usepackage{subfigure}
\usepackage{verbatim}

\begin{document}

\title{Spontaneous generation of quantum turbulence through the decay of a giant vortex in a two-dimensional superfluid}

\author{A. C. Santos}

\author{V. S. Bagnato}

\author{F. E. A. dos Santos}

\affiliation{Instituto de F\'{i}sica de S\~{a}o Carlos, Universidade de S\~{a}o Paulo, Caixa Postal 369, 13560-970 S\~{a}o Carlos, S\~{a}o Paulo, Brazil}

\date{\today}

\begin{abstract}
We show the generation of two-dimensional quantum turbulence through simulations of a giant vortex decay in a trapped Bose-Einstein condensate. While evaluating the incompressible kinetic energy spectra of the quantum fluid described by the Gross-Pitaevskii equation, a bilinear form in a log-log plot is verified. A characteristic scaling behavior for small momenta shows resemblance to the Kolmogorov $k^{-5/3}$ law, while for large momenta it reassures the universal behavior of the core-size $k^{-3}$ power-law. This indicates a mechanism of energy transportation consistent with an inverse cascade. The feasibility of the described physical system with the currently available experimental techniques to create giant vortices opens up a new route to explore quantum turbulence. 
\end{abstract}

\pacs{67.85.De, 03.75.Lm, 67.25.dk}
\maketitle

Three-dimensional (3D) turbulence in classical fluids has the energy as an inviscid conserved quantity and is characterized by the existence of an inertial range where energy is  transferred from large to small length scales through a cascade process. In the two-dimensional (2D) case, in addition to the energy, an extra inviscid quantity called enstrophy is also conserved. This turns possible the existence of two inertial ranges, one follows the Kolmogorov law $E(k)\sim k^{-5/3}$ and is associated with an inverse energy cascade; the other follows a $k^{-3}$ power law and is  related to a direct enstrophy cascade \cite{Lee1951}. 2D quantum turbulence exhibits a similar behavior as its energy spectrum also shows two regions with power law distributions, however, while the $k^{-5/3}$ region can still be associated with an inverse energy cascade, the $k^{-3}$ region is a consequence of the internal structure of singly-charged quantum vortices and there is no enstrophy cascade present \cite{Bradley2012}. Investigation of 2D quantum turbulence is an alternative to understand basic effects related to a 3D quantum turbulence recently demonstrated by our research group \cite{Henn2009b,Henn2009,Henn2009a,Seman2011a}.

In this Letter, we show the possibility of spontaneous generation of turbulence through the decay of a highly charged quantum vortex in a harmonically trapped Bose-Einstein condensate (BEC). Such a method for turbulence generation is a new and particularly promising since experimental techniques became available during the last years \cite{Mottonen2007}. Indeed, vortices with circulations as large as 60 quanta have been generated using dynamical methods \cite{Engels2003} as well as phase engineering techniques \cite{Leanhardt2002,Nakahara2000,Isoshima2007,Mottonen2002}. As we show here, it turns out that the instabilities of such giant vortices generate an effective forcing term for the incompressible part of the kinetic energy during its decay into several singly-charged vortices which constitute the main ingredient to evolution to turbulence. 

A vortex in a BEC is a manifestation of the single-valuedness of the macroscopic wave-function which describes the system. The quantization of circulation is a consequence of this topological property. In a BEC a giant vortex can be described by a wave-function of the form $\psi(\textbf{r})=f(\textbf{r})e^{i\kappa\phi}$, where a large winding number $\kappa$ indicates the existence of high angular momentum. As shown by \cite{Mottonen2007}, giant vortices decay spontaneously as time evolves through intrinsic quantum fluctuations, even though introducing a slight potential asymmetry leads to a faster growth of the unstable modes which lead to the decay, as we have verified in our simulations. 

Our system is considered to evolve according to the dissipative 2D Gross-Pitaevskii equation (dGPE)
\begin{equation}\label{gp}
i\frac{\partial\psi(\mathbf{r})}{\partial t}=(1-i\gamma)\left(-\frac{1}{2}\nabla^{2}+V(\mathbf{r})+g\left|\psi\right|^{2}-\mu\right)\psi(\mathbf{r}),
\end{equation}
where $V(\mathbf{r})=(\lambda x^2+y^2)/2$ represents the trapping potential and the anisotropy $\lambda$ is introduced in order to induce faster decay of the giant vortex. The above equation is normalized in terms of the typical scales of the trapping potential of frequency $\omega_0$, which means that time, distance, and energy, are in units of $\omega_0^{-1}$, $\sqrt{\hbar/m\omega_0}$ , and $\hbar\omega_0$, respectively. The dissipation constant $\gamma$ models the interaction of the condensate with the surrounding thermal cloud and $g$ represents the two-body interaction constant. 

In homogeneous systems (i.e., $V(\mathbf{r})=0$) the healing length is found by making the kinetic term equals the interaction in equation (\ref{gp}). Analogously, for a harmonic trapped BEC, the healing length can be defined as $\xi\equiv(gn(0))^{-1/2}$ \cite{Reeves2013}, where $n(0)=\mu/g$ is the atomic density at the center of a Thomas-Fermi cloud. In our case, we have used an interaction of $g=20000$, which makes $\xi\approx 0.11$, the Thomas-Fermi radius $r_{TF}\approx 113\xi$, and therefore $k_{TF}\equiv2\pi/r_{TF}\approx 0.50$ and $k_\xi\equiv 1/\xi\approx 8.93$, which are the wave numbers associated with the system large length and the vortex core-size, respectively. 

To simulate the decay of the giant vortex, an initial state is prepared imprinting a large circulation (i.e., winding number $\kappa$) around the center of a Thomas-Fermi cloud. By making $t\rightarrow -i t$ in equation (\ref{gp}) and setting $\gamma=0$, we shortly evolve the state for $t=0.01$ in this imaginary time description. This approximates the vortex solution to a real excited state representing a non-zero angular momentum BEC by getting rid of the density fluctuations. Although, since no stirring term is introduced to the system, a longer imaginary time evolution would leave us with a vortex-free cloud. Initially, an asymmetry of $\lambda=0.9$ is added to the harmonic potential in order to introduce a small perturbation to the initial state when evolved in real time. This procedure was done for several values of charge $\kappa$ in order to generate appropriate giant vortex initial states. In sequence, after setting $\lambda=0$, a real time evolution is performed for $t=200$ with a fixed dissipation constant of $\gamma=0.001$. 

All numeric simulations were performed for a domain of $-20\leq x \leq 20$ in a grid of $1024\times 1024$ using the adaptive Runge-Kutta methods in Fourier space with the help of XMDS \cite{Dennis2013}.  

To extract the kinetic energy spectra from our simulations, it is useful to define a density weighted velocity field of the form $\textbf{w}(\textbf{r},t)\equiv \left|\psi(\textbf{r},t)\right|\textbf{v}(\textbf{r},t)$. It is then possible to decompose this vector field into the sum $\textbf{w}(\textbf{r},t)=\textbf{w}_c(\textbf{r},t)+\textbf{w}_i(\textbf{r},t)$, whose terms satisfy $\nabla\times\textbf{w}_c=0$ and $\nabla\cdot\textbf{w}_i=0$. The kinetic energy is therefore split into compressible and incompressible parts, $E_c$ and $E_i$, respectively related to the sound field and vortices \cite{Bradley2012}.

\begin{figure}[t]
\begin{center}
\includegraphics[width=0.5\textwidth]{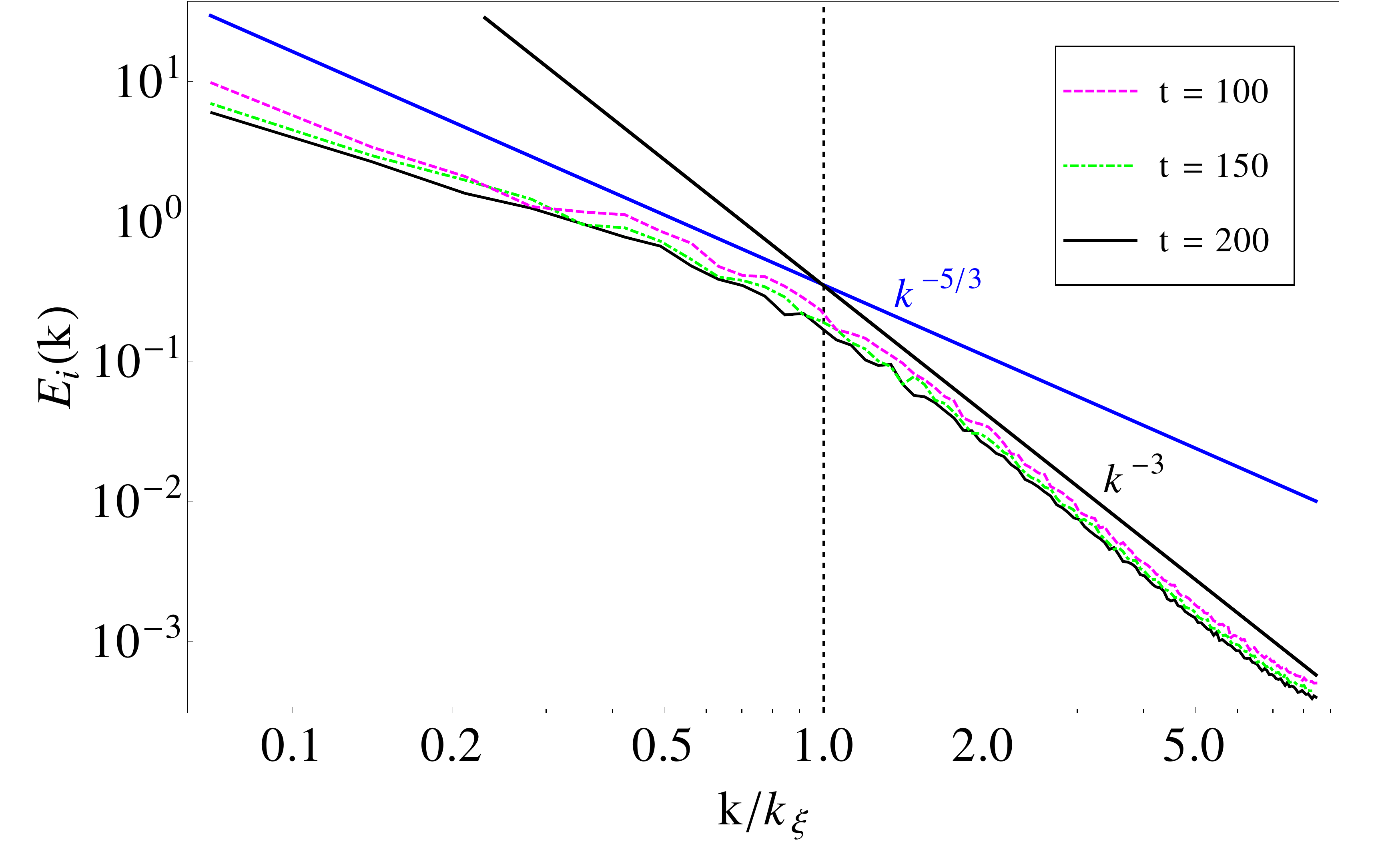} 
\caption{(Color online) Incompressible kinetic energy spectra for a 40-charged vortex decay at different instants of time evolution.}
\label{k40p1}
\end{center}
\end{figure}

Simulations were performed for charges ranging from $\kappa=2$ up to $\kappa=60$. The same features described here for the particular case of $\kappa=40$ were verified for charges as low as $\kappa=10$. The spectra taken from the simulation of a 40-charged vortex decay presented in FIG. \ref{k40p1} illustrate well our results.  

\begin{figure}[h]
  \centering
  \subfigure[Giant vortex initial state at $t=0$.]{\includegraphics[width=0.23\textwidth]{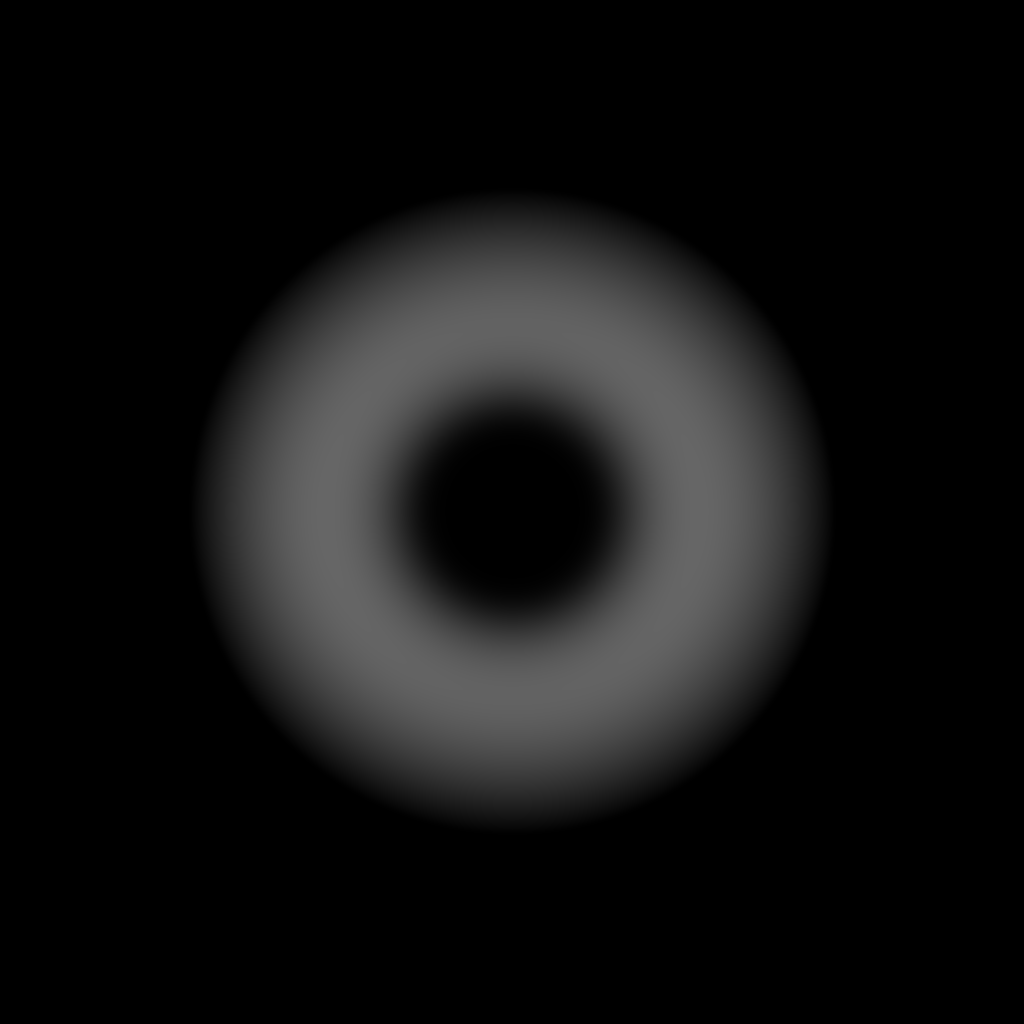}}
  \subfigure[Vortex decay at $t=17$.]{\includegraphics[width=0.23\textwidth]{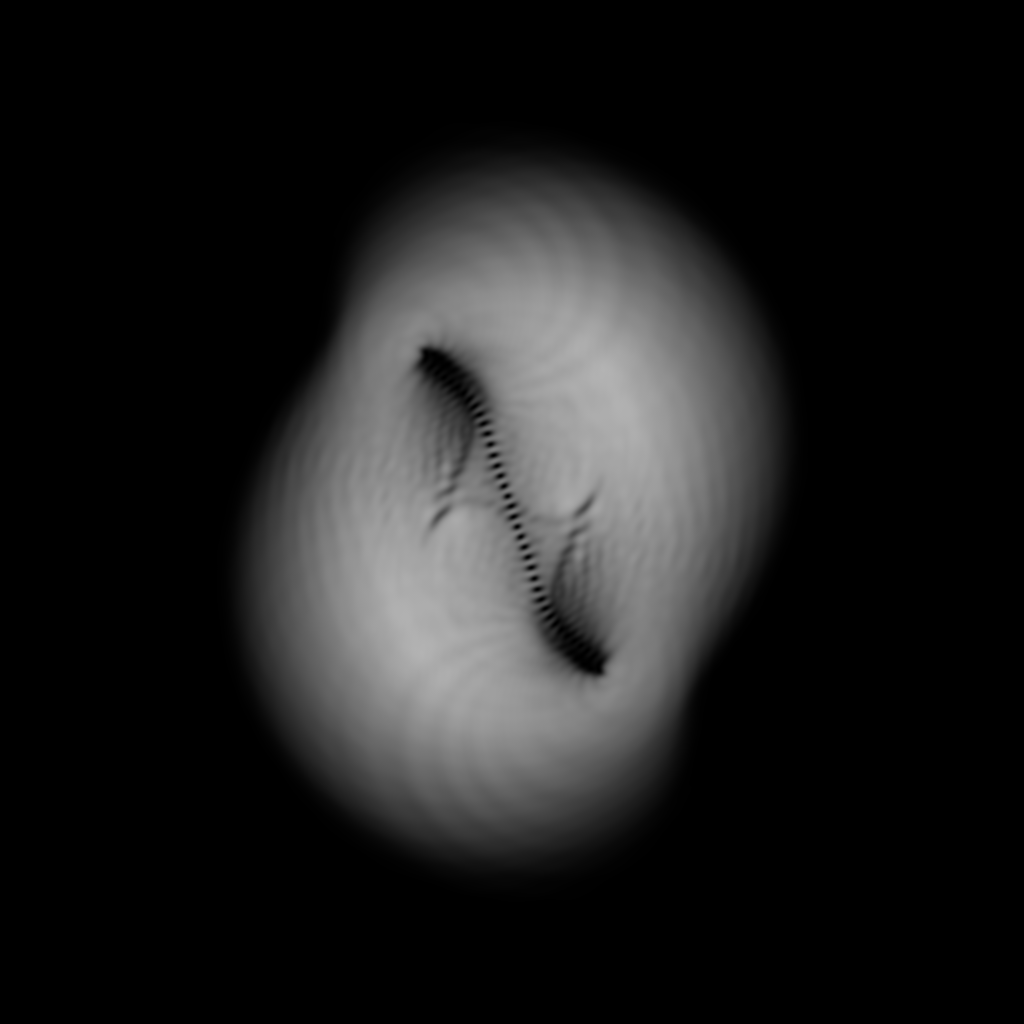}}
  \subfigure[Clusters formation at $t=20$.]{\includegraphics[width=0.23\textwidth]{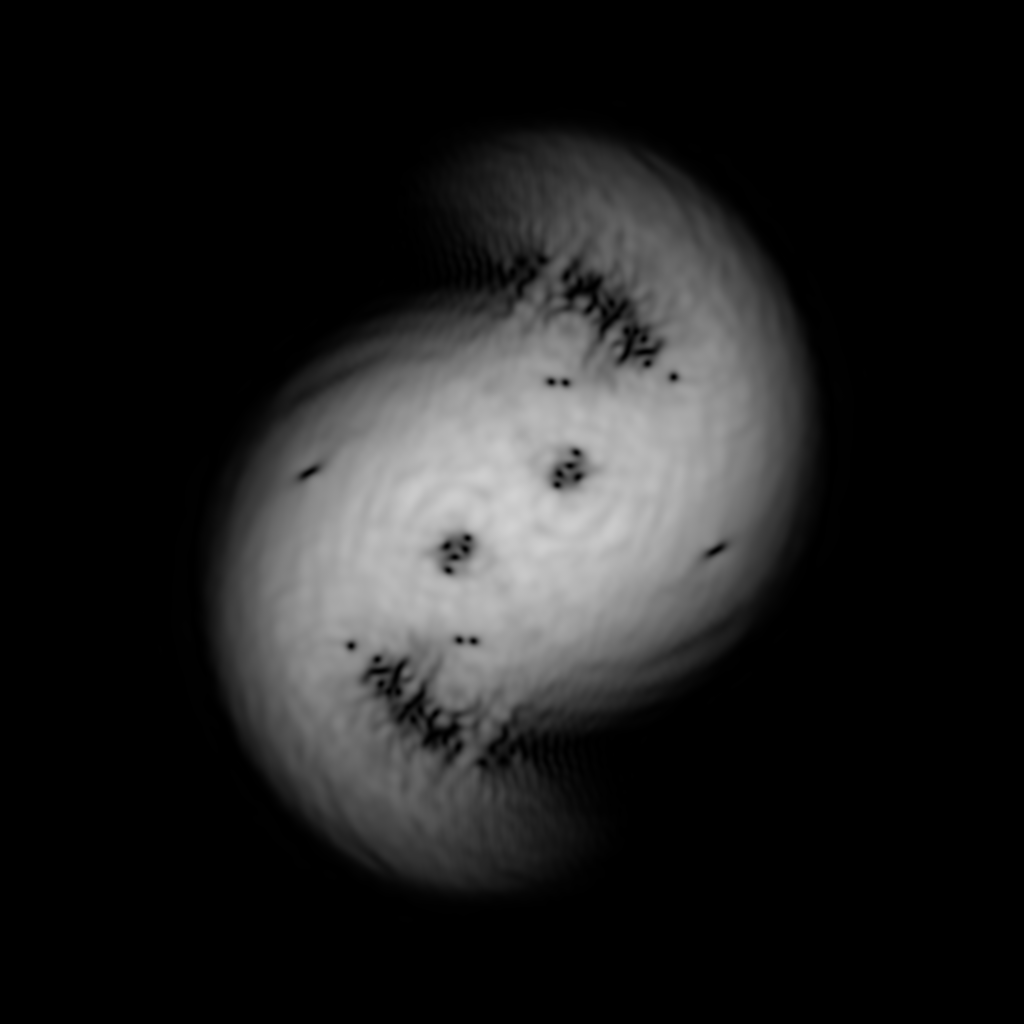}}
  \subfigure[Additional clusters formed at $t=30$.]{\includegraphics[width=0.23\textwidth]{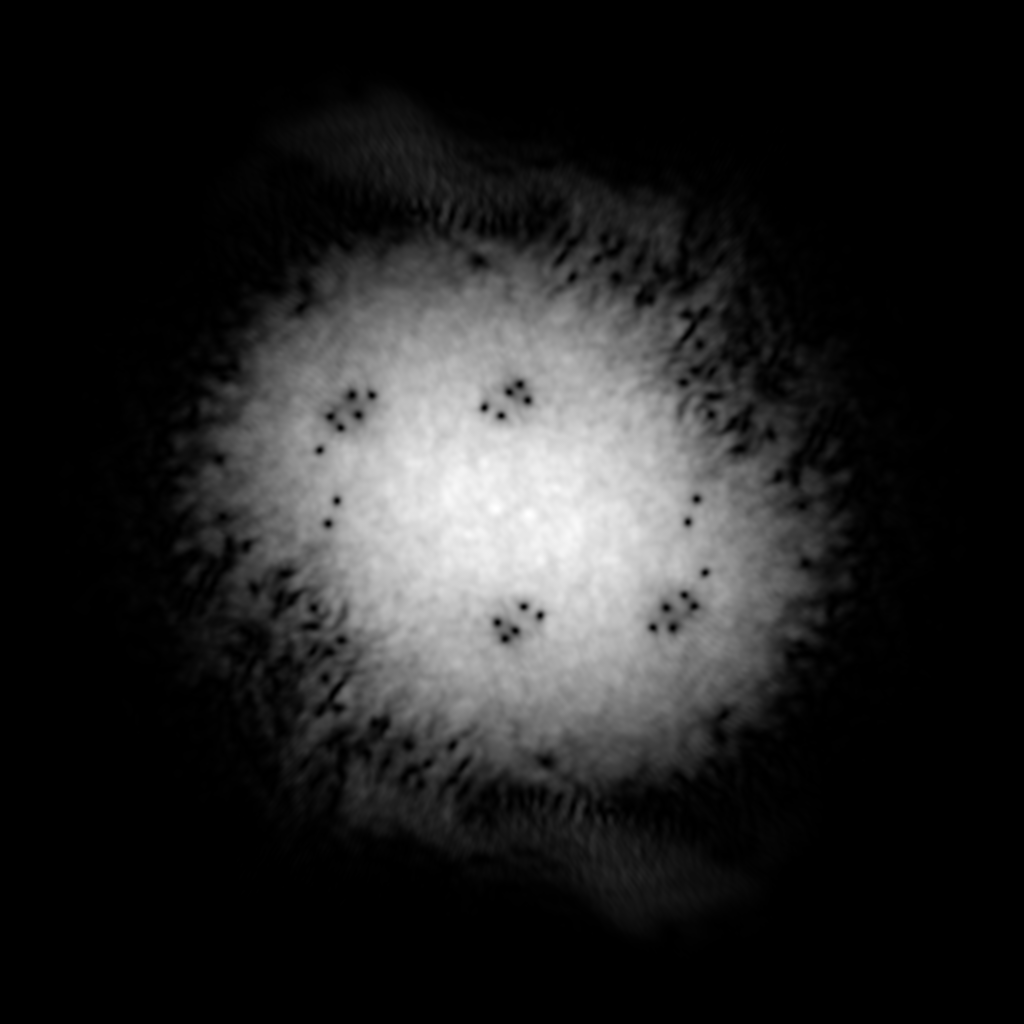}}
  \subfigure[Merging of clusters at $t=35$.]{\includegraphics[width=0.23\textwidth]{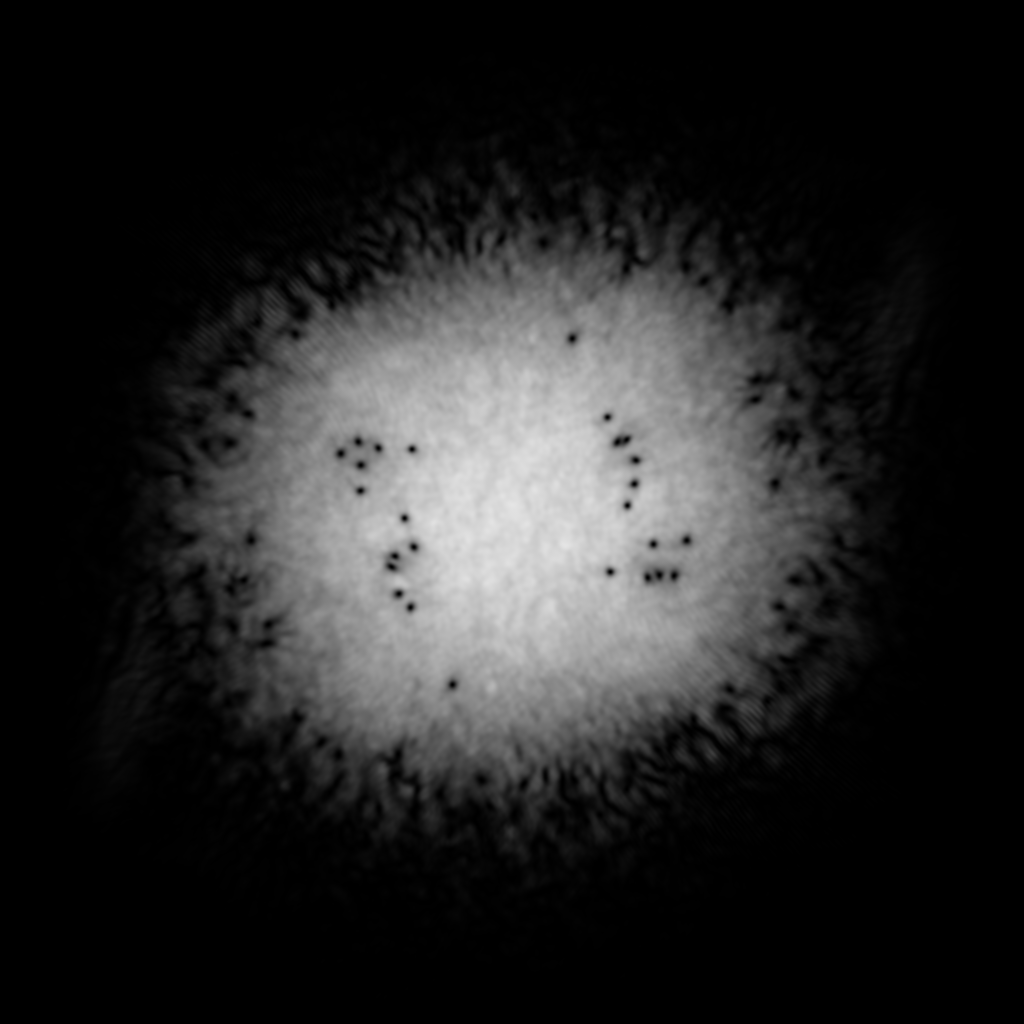}}  
  \subfigure[Final single-charged vortex distribution at $t=200$.]{\includegraphics[width=0.23\textwidth]{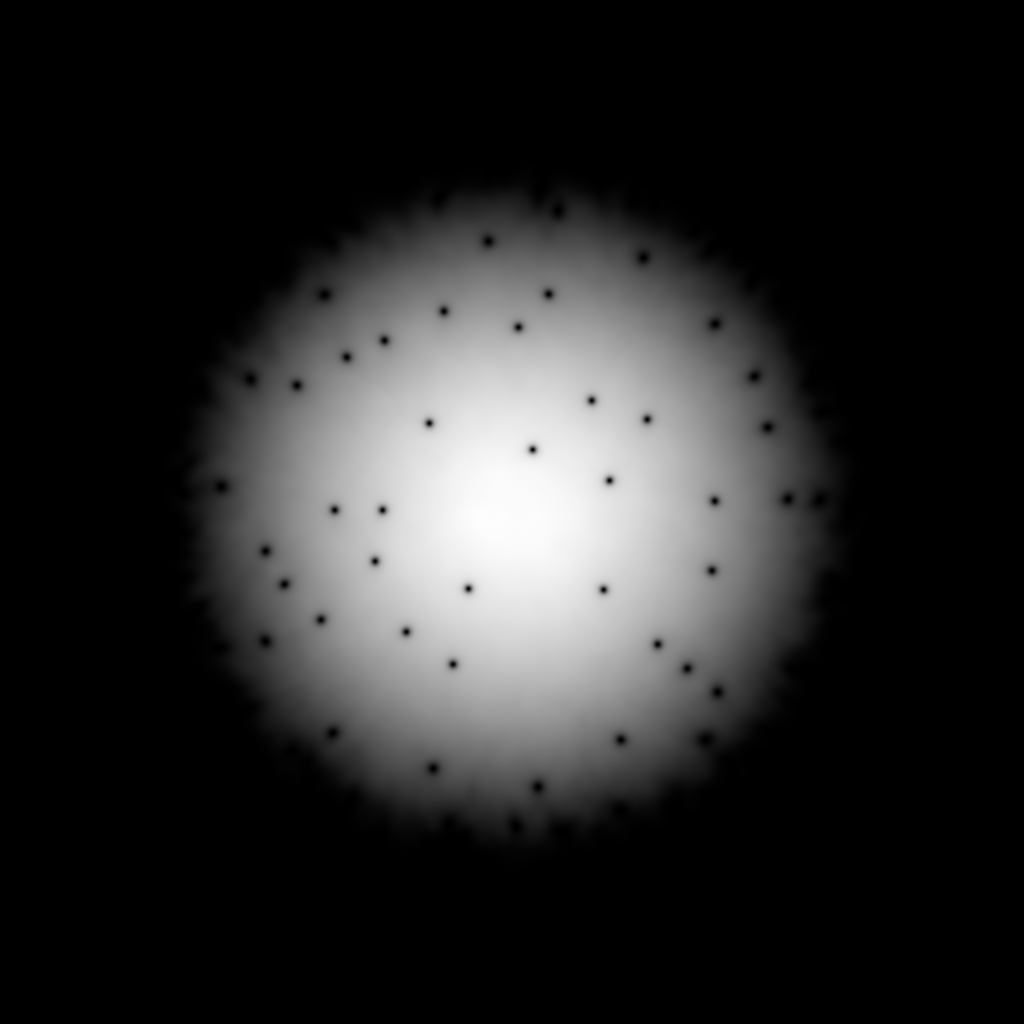}}
  \caption{Time evolution of the 40-charged vortex decay. Clustering as an evidence of inverse energy cascade.}
  \label{clustering} 
\end{figure}

\begin{figure}[t]
\begin{center}
\includegraphics[width=0.5\textwidth]{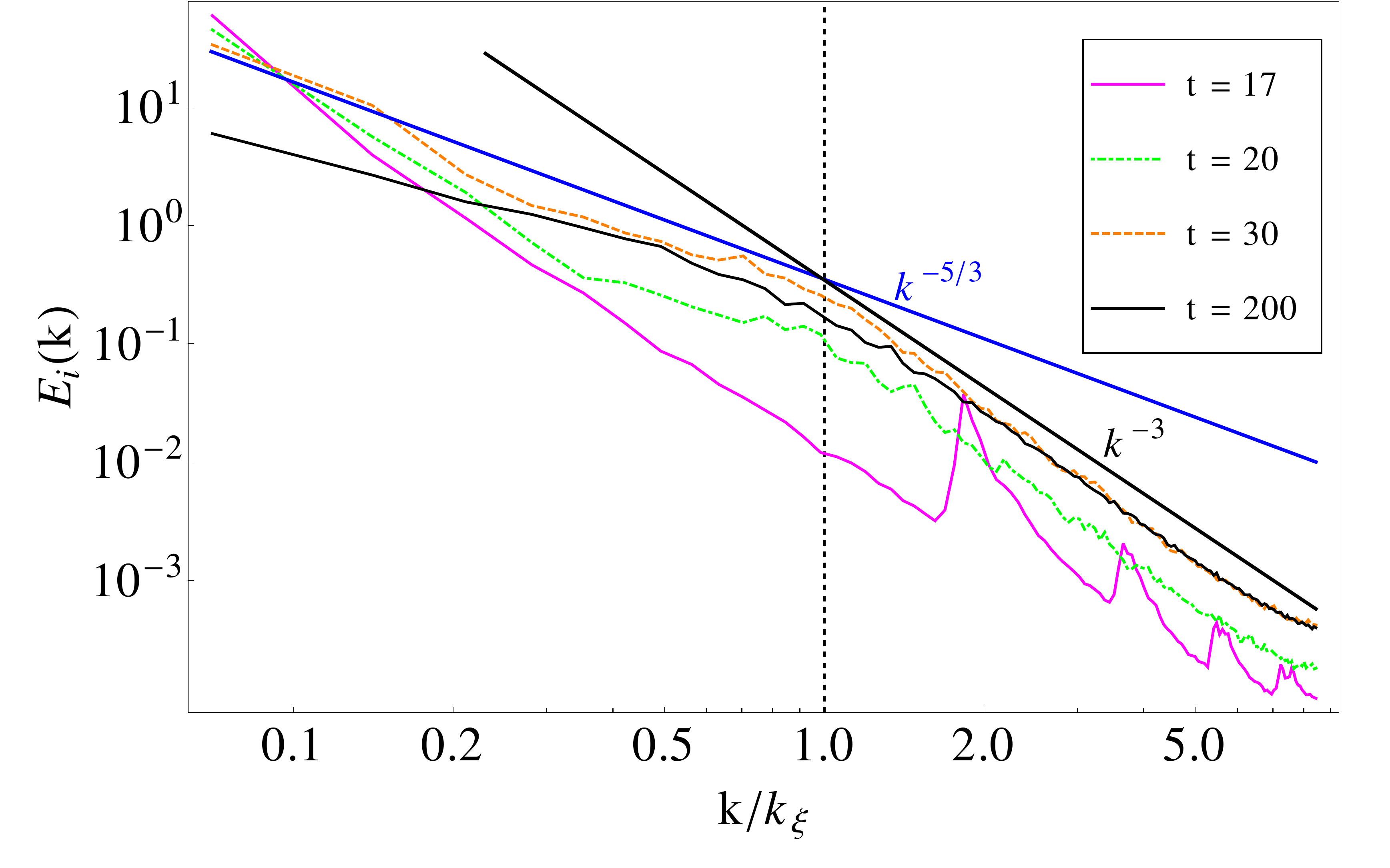} 
\caption{(Color online) Incompressible kinetic energy injection for large k through the process of 40-charged vortex decay at different instants of time evolution.}
\label{peak}
\end{center}
\end{figure}

\begin{figure}[h]
  \centering
    \subfigure[Compressible kinetic energy.\label{subfigure label2}]{\includegraphics[width=0.48\textwidth]{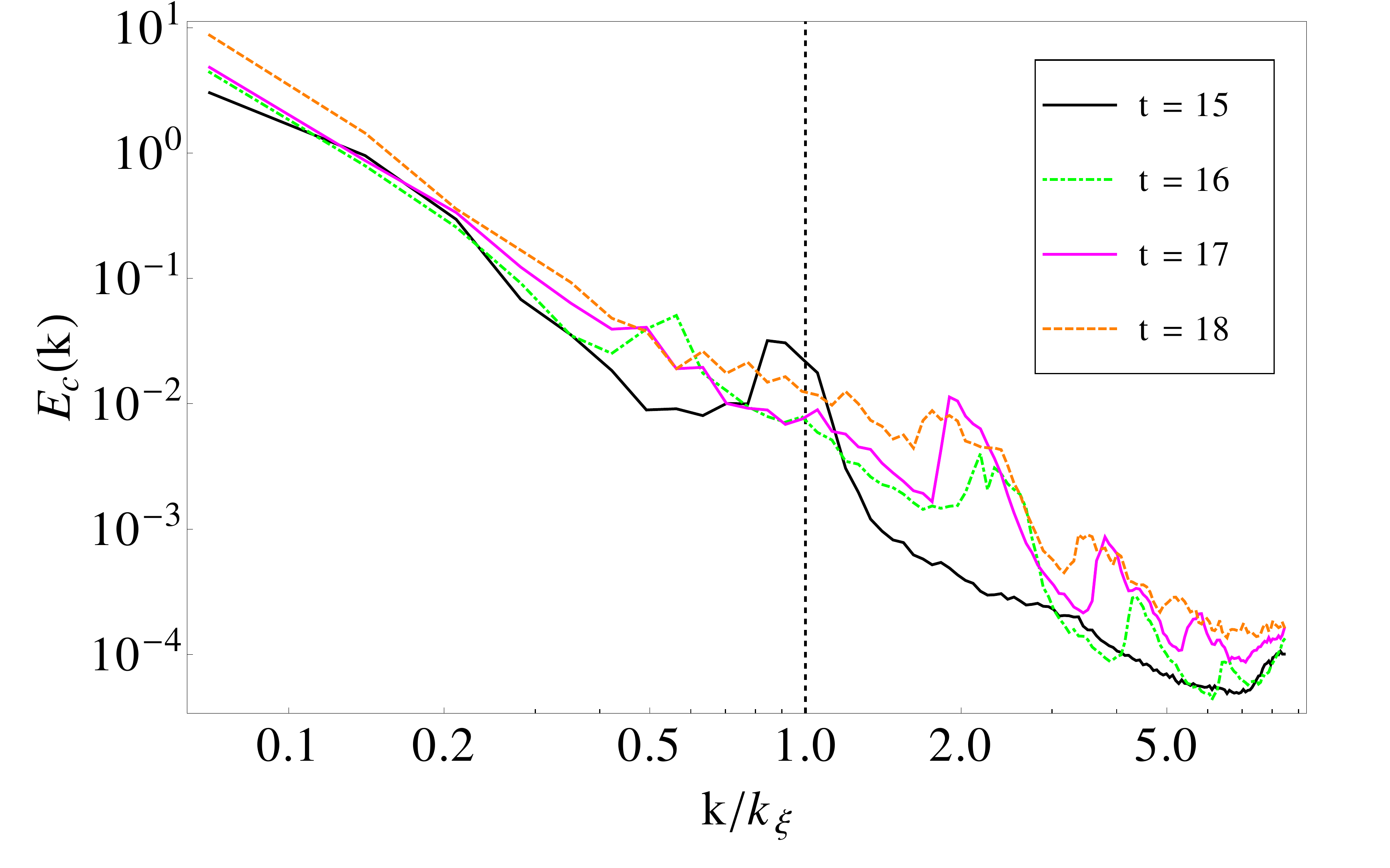}}
    \subfigure[Incompressible kinetic energy.\label{subfigure label}]{\includegraphics[width=0.48\textwidth]{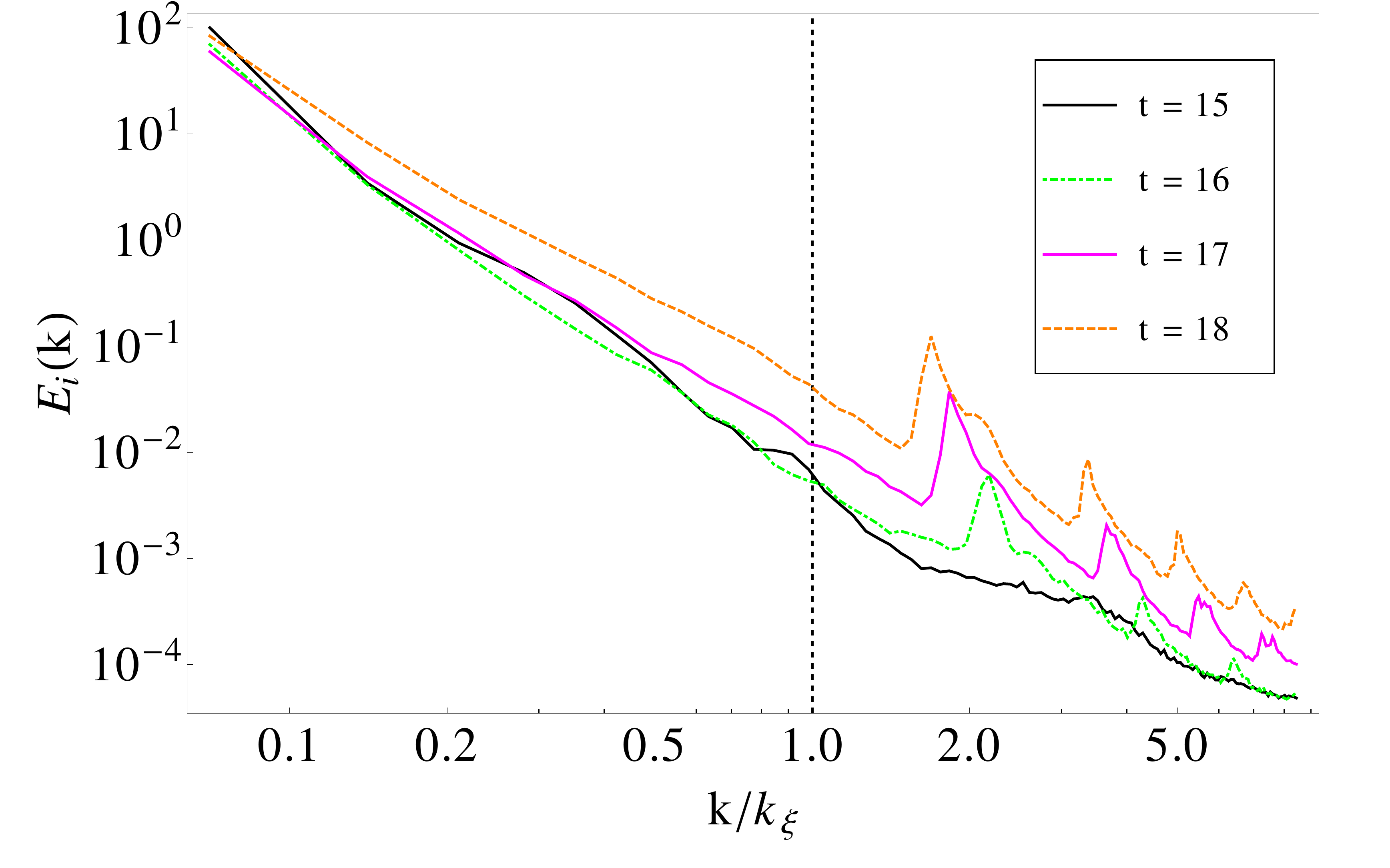}}
  \caption{(Color online) Coupling between incompressible and compressible part of the velocity field. A peak of energy appears first for the compressible part at time $t=15$, which is not verified in the incompressible part.} 
  \label{coupling}
\end{figure}

From the time sequence shown in FIG. \ref{clustering} we verify the important feature of 2D quantum turbulence, which is the formation, merging, and enlargement of clustered structures. This is visually endorsed by images (c), (d), and (e), where clusters of same-signed vortices expand as time evolves. This phenomenon is usually associated with an inverse cascade, in which injection of energy happens for small length scales and is transferred to larger scales, in this case some orders of $\xi$. This can also be substantiated by analyzing the incompressible kinetic energy spectra shown in FIG. \ref{peak}. As soon as the giant vortex decays at instant $t=17$, we notice the presence of peaks, associated to the injection of energy into those specific scales. This is analogous to the forcing reported in \cite{Reeves2013} and supports the idea of an internal forcing term, in the sense that it is generated by the own system metastability, while energy is injected merely through the giant vortex decay. Each peak may be associated to a Bogoliubov mode excited by the perturbations on the system which lead to the vortex decay, being a result of particular unstable modes growth \cite{Kuopanportti2010}. As time evolves,  a raise of the incompressible energy for a range of $0.2 \lesssim k/k_{\xi} \lesssim 1.0$ can be verified, suggesting a flux from the large to small $k$. Furthermore, the lowering of the spectra for low values of momenta ($k/k_{\xi} \lesssim 0.1$) indicates dissipation acting more effectively on that region. 

The nature of the forcing term can be understood in terms of the comparison between compressible and incompressible kinetic energy spectra for instants close to the decay. As shown in FIG. \ref{coupling}, there is an injection of energy at time $t=15$ for $k/k_{\xi}\approx 0.9$, which is not verified in the incompressible part for the same instant. Analogously to what was verified in \cite{Reeves2012}, this peak can be attributed to a compression wave formed due to the release of sound-like energy in the moment of the decay. In the subsequent time, $t=16$, both incompressible and compressible spectra show peaks formation, which can only be understood as the coupling between the two energies, explaining the mechanism of energy injection into the superfluid incompressible part.   

In the context of  2D forced quantum turbulence, reference \cite{Reeves2012} used a gaussian paddle to generate vortices and anti-vortices in the system and found similar energy spectra compared to our results. Both, \cite{Reeves2012} and our present results show the bilinear log-log plots power-laws with close resemblance to the Kolmogorov $k^{-5/3}$ law. We also found for the infrared part of the spectrum a small deviation as verified for different scenarios of forcing in \cite{Reeves2012,Reeves2013}. In the latter, the authors justify this behavior by assuming that the size of the paddle (i.e., typical width of the gaussian potential used to stir the condensate) becomes an impediment for the generation of larger clusters of same-signed vortices. In our case, since we do not have a constant forcing structure, it is believed that the system finiteness is an intrinsic inhibition to larger structures development. Also, since the nature of the forcing in the system is a fast decay of a multicharged vortex, the sudden ceasing of this localized forcing may contribute to an incomplete evolution of the turbulent state. 

We notice that the ultraviolet portion ($k>k_{\xi}\equiv \xi^{-1}$) of the incompressible energy spectra shows a robust power law of $k^{-3}$  \footnote{ Which in \cite{Reeves2012} is referred as \textit{onstrophy} power-law to emphasize to the role of the Onsager vortex quantization.} that is associated to the quantization of vortices, therefore being an essential characteristic of the 2D quantum turbulence. Several works have shown this feature in different contexts of forced turbulence and has been generically discussed in \cite{Bradley2012}.

The effect of the system finiteness can be in principle diminished by tuning the interaction term $g$ in the dGPE in order to increase the size of the BEC cloud. But these latter systems would demand larger multicharged vortices to reproduce similar results, which is still a challenge with the current procedures. Nevertheless, vortex pumping and artificial gauge fields are promising experimental techniques to explore the phenomenon described here. 

As the charge was increased, the stationary state became more definite, with less oscillations in the quasi-Kolmogorov region of the spectra. The universality of the power-law $k^{-3}$ was always verified, in accordance to previous works on 2D quantum turbulence.

We have shown the generation of 2D quantum turbulence by simulating the decay of a giant vortex whose evolution is described by the dGPE. Elements such as the bilinear log-log plot of the incompressible kinetic energy, alongside the formation, merging, and enlargement of vortex clusters, indicate the inverse energy cascade, also present in 2D classical turbulence. In a direct analogy, the decay of a giant vortex in a three-dimensional system may induce reconnection of smaller vortex lines, generating the characteristic 3D quantum turbulence spectra. Therefore, this work opens a new route to explore the phenomenon of quantum turbulence, while assuring the experimental feasibility of the described system and allowing a high controllability of initial conditions. Extension to a 3D system with spontaneous generation of turbulence is also a possibility presently under consideration.

We have special thanks to Professor Jos\'{e} Abel Hoyos Neto for important suggestions and support. This research was financially supported by FAPESP, and CNPq. The N\'{u}cleo de Apoio a \'{O}ptica e Fot\^{o}nica (NAPOF-USP) is acknowledged for computational resources.

\bibliography{bib}

\end{document}